\documentclass{article}
\usepackage{mrs2005,epsfig}
\begin{document} 
\title{CONFRONTING HIERARCHICAL CLUSTERING MODELS WITH OBSERVATIONS OF GALAXY 
PAIRS}

\author{J. P\'EREZ$^{2,1}$, P.B. TISSERA$^1$, D. G. LAMBAS$^{3}$, C. SCANNAPIECO$^1$}
\affil{$^1$ Instituto de Astronom\'{\i}a y F\'{\i}sica del Espacio (IAFE), Casilla de Correos 67,
Suc. 28, 1428, Buenos Aires, Argentina.\\
$^2$ FCAGLP-IALP, Paseo del Bosque S/N, La Plata, Buenos Aires, Argentina\\
$^3$ IATE, Observatorio Astron\'omico de C\'ordoba, Laprida 922, C\'ordoba, Argentina}

\begin{abstract} 
We investigate the star formation activity  in galaxy pairs in 
chemical hydrodynamical simulations consistent with a $\Lambda$-CDM scenario.
A statistical analysis of the effects of  galaxy interactions 
on the star formation activity as a function of orbital parameters shows that 
close encounters ($r <30$ kpc $h^{-1}$) can be effectively correlated with  an  enhancement of  
star formation activity with respect to galaxies without a close companion.
Our results suggest  that the stability properties of systems are also relevant in this process.
We  found that the passive  star forming galaxies  pairs
tend to have deeper potential wells, older stellar populations, 
and less leftover gas than active star forming ones.
In order to assess the effects that projection and interlopers could introduce
in observational samples, we have also constructed and analysed projected 
simulated catalogs of galaxy pairs. In good agreement with  observations,
our results show a threshold  ($r_{p} <25$ kpc $h^{-1}$) for interactions
to enhance the star formation activity with respect to galaxies without
 a close companion. Finally, analysing the environmental effect, 
we  detect the expected SFR-local density
relation for both pairs and isolated galaxy samples, although the density 
dependence is stronger for galaxies in pairs suggesting a relevant role
for interactions in driving this relation.

\end{abstract} 
 
\section{Introduction} 
 
Observations show that mergers and interactions   can  induce star 
formation (SF) activity  in galaxies (e.g. Larson and Tinsley 1978).
Barton et al. (2000) and Lambas et al. (2003, LTAC03) analysed a sample
of pair galaxies in the field finding a clear correlation between  the proximity in
projected distance and radial velocity of two galaxies and their SF activity.
Recently, Alonso et al. (2004) claimed that the mechanism for triggering SF in 
interactions seems to be  weakly dependent on environment, finding that even in cluster
environment, the relation between proximity and SF activity has very similar characterisicts
to the one measured in the field.
On the other hand, numerical simulations of pre-prepared mergers showed that 
interactions between axisymmetrical systems without bulges might induce gas inflows 
to the central region of the systems, triggering starburst episodes 
(Mihos  and Hernquist 1996).
Results from the study of the effects of mergers in
the SF history of galactic objects in  cosmological hydrodynamical simulations (Tissera et al.
2002),
indicate
that, during some mergers events, gaseous disks could experience
two starbursts depending on  the characteristic of the potential well. Other authors showed that
orbital parameters of the encounter can also play a role in the triggering of SF activity
(e.g. Barnes \& Hernquist 1996).

Undoubtely, galaxy-galaxy  interactions are a key process in the regulation of star formation 
activity and if the Universe is consistent with a hierarchical scenario, then it is of
utmost importance to understand the detailed characteristics, effects and efficiency of the action  of
this physical process on the life of galaxies.
In this work we focus on a stasticial analysis  of galaxy pairs  in a
hierarchical scenario  with the aim at confronting this scenario with 
recent observational results of galaxy pairs. Details can be found in 
P\'erez et al. (2005).

\section{Results and Conclusions} 

We analyzed a 10 Mpc $h^{-1}$ cubic volume of   a $\Lambda$-CDM Universe
($\Lambda=0.7$, $\Omega=0.3$, $H=100 h$ ${\rm km s^{-1} Mpc^{-1}}$ with $ h=0.7$ ) run with  the chemical cosmological GADGET-2 (Scannapieco 
et al. 2005). The chemical GADGET-2 uses a SF algorithm based on the Schmidt law (Navarro \&
White 1994), but  transforming gas into stars in a stochastic way. It also 
includes chemical evolution by describing  the enrichment of the interstellar medium by SNII and SNIa.
After identifying the bounded systems in the simulations and impossing
a minimun stellar mass of $8\times 10^8 M_{\odot} h^{-1} $, 
we constructed a tridimensional  simulated galaxy  catalog (3D-GP) selecting pairs
according to a distance  criterium. Analysing the dependence of the SF 
activity on the proximity to a near companion, we found a sharp increase of the SF in 
galaxies within  $r<100$ kpc $h^{-1}$, which defines a suitable threshold to 
select pairs from the 3D distribution.
We also build a projected galaxy pair (2D-GP) catalog by projecting the total 3D galaxy distribution
onto random directions and then, selecting   2D pairs  according to the 
observational criteria of Lambas et al. (2003). Hence, the 2D-GP catalog is
formed by those systems with  relative projected 
separation  $r_{p}<100$ kpc $h^{-1}$  and radial velocity  $\Delta cz <350$ km s$^{-1}$. 
In order to provide a suitable comparison to underpin the effects of 
interactions,  we constructed  galaxy control samples for  pair catalogs defined 
by galaxies without a close companion within the corresponding thresholds.
For each simulated galaxy, we estimated the stellar birthrate parameter
$b$, defined as the present level of SF activity of a galaxy
normalized to its mean past SF rate.

The analysis of the SF activity for galaxies in the 3D-GP catalog as a function 
of their orbital parameters shows that proximity can be statistically related to an increase
of SF activity,   at higher 
levels than those measured for galactic systems without a close companion, 
if systems are closer than $\approx 30\pm 10$ kpc $h^{-1}$.  
On the other hand, we find a very weak indication for encounters with low relative velocities 
to be related with an enhancement of SF activity. 
Interestingly, we also detect  that the triggering of SF by tidal interactions
can be statistically related to   the stability of the systems
and  the gas reservoir. In fact, we found that $\approx 56\%$ of galaxies in close pairs
  ($r < 30 $ kpc $h^{-1}$) are forming
stars at  lower level than the average of the control sample. Part of these systems
have experienced recent star formation activity and the rest shows deeper
potential well and are gas poorer than galaxies in pairs with strong SF activity.   

Results from the 2D-GP catalog 
show the same global trends detected in the 3D-GP sample.
The enhancement threshold in projected distance drops with
respect to that found in 3D to  $\sim 25 \pm 5$ kpc $h^{-1}$ which is in good agreement
with observational results (LTAG03). 
This shrinking in the threshold is produced by both geometrical projection effects 
and interlopers (spurious pairs). In order to separate these both effects, we have removed 
spurious pairs from the  2D-GP sample by checking their 3D relative separations. Consistently with
previous works (Alonso et al. 2004; Mammon et al. 1997), we found that $30\%$ of 
the pairs  in  2D-GP sample 
are spurious. This percentage reduces to    $19\%$ for  2D close pairs 
($r_{p}<25 $ kpc $h^{-1}$ and $\Delta cz<100 $ km s$^{-1}$). 

On the other hand, we analysed the dependence of the SF on environment
by  calculating the projected local density parameter: 
$\Sigma=6/(\pi d^{2}_{6})$, being $d_{6}$ the projected distance to the $6^{th}$ 
neighbor brighter than $M_{r}=-20.5$. Globally, the SF activity in pairs
seem to depend weakly on environment. However, the fraction of star forming
pairs is higher in low density regions.

Summing up, from the 3D-simulated galaxy pair catalog, we found that galaxy-galaxy interactions can be
correlated with  an enhacement of SF activity  at higher levels 
than those measured for galactic systems without a close companion.
We also found that the internal dynamical stability of galactic systems plays an important role 
as it can be deduced from the presence of an anticorrelation signal
 between the deepness of the potential well and the 
star formation activity.
From the  analysis of the projected galaxy pair catalog  we estimated similar dependence to
those observed in recent observational works. The fraction of spurious pairs is found
to be higher for larger galaxy pair separations and to increase with local density. However,
the trends estimated from the projected catalog are in good agreement to those calculated
from the tridimensional one.
We also found that the star formation activity in galaxies in pairs is weaky dependent on
the cosmology and local environment, although the fraction of strong star forming galaxies in
pairs increases with decreasing density in agreement with observations.

\acknowledgements{This work was partially supported by the Consejo Nacional 
de Investigaciones Cient\'{\i}ficas y T\'ecnicas and Fundaci\'on Antorchas. Simulations
were run on Ingeld PC Cluster funded by Fundaci\'on Antorchas. JP and CS 
thank the LOC of this meeting for their help which made their participation possible.
}

\vfill 
\end{document}